\documentclass[11pt]{article}
\pdfoutput=1

\usepackage{amsmath}
\usepackage{amssymb}
\usepackage{graphicx}
\usepackage{cite}
\usepackage[margin=1in]{geometry}
\usepackage[draft=false]{hyperref}
\usepackage[labelfont=bf,labelsep=period,justification=raggedright]{caption}
\usepackage[justification=raggedright,singlelinecheck=off,labelformat=simple,labelfont=bf]{subcaption}

\usepackage[caption2]{ccaption}
\usepackage{rotating}

\date{\today}

\begin{document}

\begin{flushleft}
{\huge
\textbf{Transformation of stimulus correlations by the retina}
}
\\\vspace{3ex}
{\large
Kristina D.\ Simmons$^{1,\dagger}$, 
Jason S.\ Prentice$^{2,6, \dagger}$, 
Ga\v{s}per Tka\v{c}ik$^{3}$,
Jan Homann$^{2}$, 
Heather K.\ Yee $^{4}$, 
Stephanie E.\ Palmer $^{4}$,
Philip C.\ Nelson$^{2}$, 
Vijay Balasubramanian$^{1,2,5,\ast}$
}
\\\vspace{2ex}
{\small
\bf{1} Department of Neuroscience, University of Pennsylvania, Philadelphia PA 19104, USA
\\
\bf{2} Department of Physics, University of Pennsylvania, Philadelphia PA 19104, USA
\\
\bf{3} Institute of Science and Technology Austria, A 3400 Klosterneuburg, Austria
\\
\bf{4} Department of Organismal Biology and Anatomy, University of Chicago, Chicago IL 60637, USA
\\
\bf{5} Laboratoire de Physique Th\'{e}orique, \'{E}cole Normale Sup\'{e}rieure, 75005 Paris, France
\\
\bf{6} Princeton Neuroscience Institute, Princeton University, Princeton NJ 08544
\\
$\ast$ E-mail: vijay@physics.upenn.edu}
\\\vspace{1ex}
$\dagger$ KDS and JSP contributed equally to this work.
\end{flushleft}

\section*{Abstract}
Redundancies and correlations in the responses of sensory neurons seem to waste neural resources but can carry cues about structured stimuli and may help the brain to correct for response errors. To assess how the retina negotiates this tradeoff, we measured simultaneous responses from populations of ganglion cells presented with natural and artificial stimuli that varied greatly in correlation structure.  We found that pairwise correlations in the retinal output remained similar across stimuli with widely different spatio-temporal correlations including white noise and natural movies.   Meanwhile, purely spatial correlations tended to increase correlations in the retinal response.   Responding to more correlated stimuli, ganglion cells had faster temporal kernels and tended to have stronger surrounds.  These properties of individual cells, along with gain changes that opposed changes in effective contrast at the ganglion cell input, largely explained the similarity of pairwise correlations across  stimuli where receptive field measurements were possible.

\section*{Introduction}
An influential theory of early sensory processing argues that sensory circuits should conserve scarce resources in their outputs by removing correlations present in their inputs \cite{barlow61, srinivasan82, atick92}.  However, recent work has clarified that some redundancy in the retinal output is useful for hedging against noise \cite{borghuis08, tkacik10}.  Moreover, sensory outputs with varying amounts of correlation can engage cortical circuits differently and thus result in a different sensory ``code'' \cite{estebanez12}.  Thus, some degree of redundancy appears to be useful to the brain when dealing with response variability and making decisions based on probabilistic input \cite{barlow01}.     Indeed, correlations between neurons in visual cortex are largely unchanged between unstructured and naturalistic visual stimuli \cite{fiser04}.   Thus we hypothesized that {\it retina may adjust to the spatio-temporal structure of stimuli not to decorrelate but to maintain a relatively invariant degree of output correlation}.     Previous studies have examined pairwise correlations amongst retinal ganglion cell spike trains in specific stimulus conditions \cite{ganmor11, puchalla05, schneidman03, trong08, greschner11, pitkow12} but did not study the changes in correlation for the same pairs across stimuli.  

Are there mechanisms that might allow the retina to adjust its functional properties when stimulus correlations change?  Traditionally, retinal ganglion cells (RGCs) have been described by a fixed linear receptive field followed by a static nonlinearity \cite{rodieck65}, where surround inhibition acts linearly to suppress pairwise correlations in natural visual input \cite{srinivasan82, atick92}.  
In this view, the receptive field and nonlinearities might vary dynamically with stimulus correlations, possibly by changing the strength of lateral inhibition to maintain a fixed amount of output correlation.   Indeed, correlation-induced changes in receptive fields have been observed in the LGN and visual cortex \cite{lesica07, sharpee06}.

To test these ideas, we performed a series of experiments in which we presented the retina with several stimuli with varying degrees of spatial and temporal correlations.   The retina never fully decorrelated its input; even for white noise stimuli some correlations were present between pairs of retinal ganglion cell spike trains. Responding to natural movies, however, output correlations were only slightly greater than they were while responding to white noise, despite the dramatic difference in input correlations.  We found a similar result for a spatio-temporal exponentially correlated stimuli, with the increase in output correlations being smaller still. For stimuli with high spatial, but not temporal correlations, output correlation increased with input correlation to a larger degree than in natural movies. Thus, pairwise output correlations are similar over a broad range of spatio-temporal correlations but increase with spatial correlation in the absence of temporal correlation.  Additionally, we observed a faster response timecourse and a skew towards stronger inhibitory surrounds in response to correlated stimuli. These changes were sufficient to largely explain the observed suppression of pairwise correlations in the retinal output.

\section*{Results}

\subsection*{Simultaneous measurements of ganglion cell responses}
\begin{figure}
\begin{center}
\includegraphics[keepaspectratio, width=\linewidth]{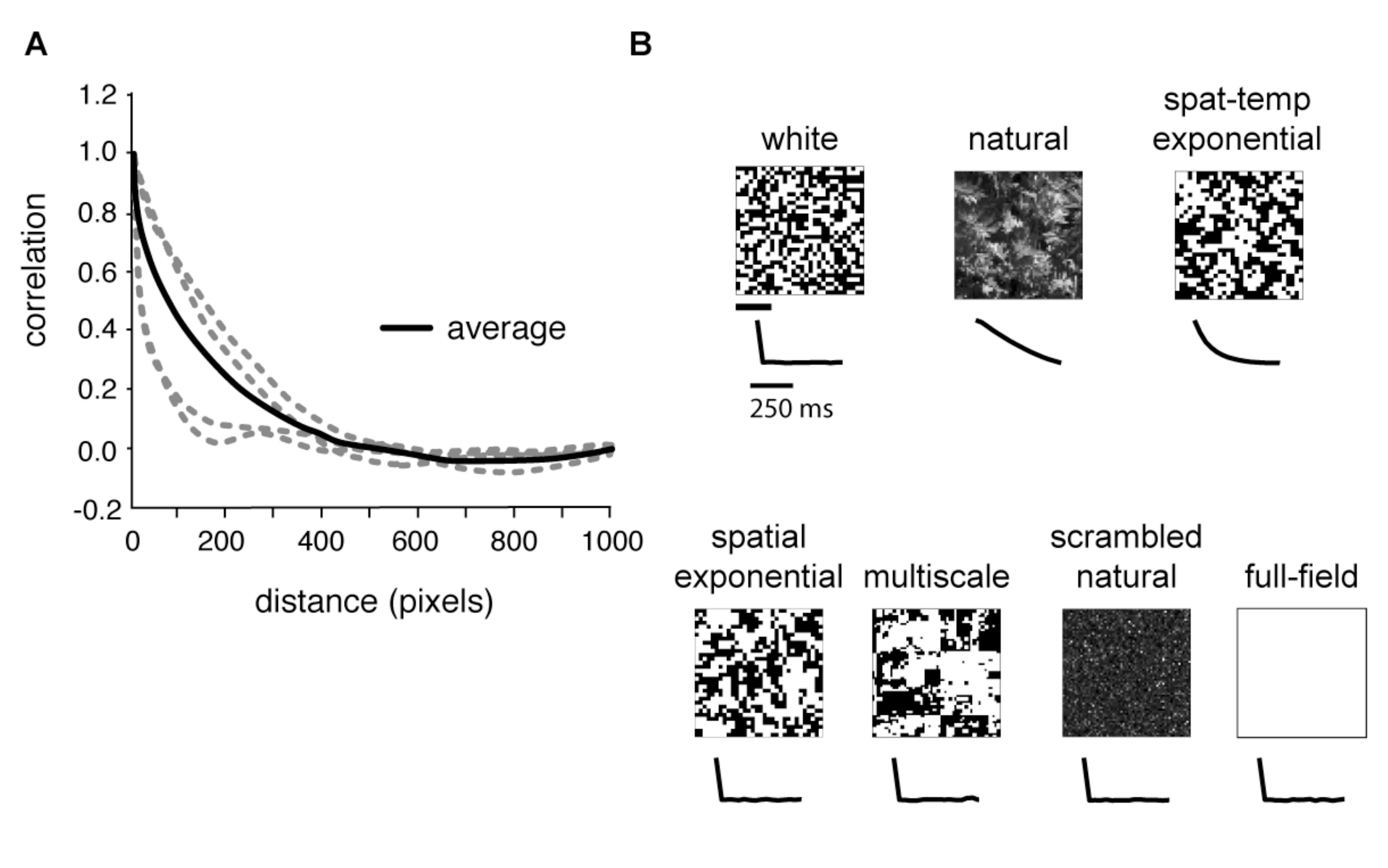}
\phantomsubcaption\label{f:natimages}
\phantomsubcaption\label{f:stim}
\end{center}
\caption{
{\bf Natural and artificial stimuli vary in correlation structure.}  
\textbf{(A)} Spatial correlation functions from four natural images, in gray.  Black line shows average correlation function over a large database of natural images.  Although all images' correlation functions have the same general shape, there are clear differences between images.
\textbf{(B)} Examples of the stimuli used in this work.  Traces above frames show the temporal correlation function of each stimulus. Stimuli were displayed at 30 Hz in alternating 10-minute blocks. Spatial scale bar is $400\ \mathrm{\mu m}$. 
}
\label{Fig1}
\end{figure}

	We used a multi-electrode array to measure simultaneous responses from groups of $\sim$40 retinal ganglion cells in guinea pig. Each recording interleaved 10-minute blocks of white noise checkerboard stimuli with 10-minute blocks of correlated stimuli.  Example frames from each stimulus are shown in Fig.\ \ref{f:stim}, together with their respective temporal correlation functions.  We probed retinal responses to natural movies, which allowed us to determine properties of ganglion cell population activity during natural vision.  However, natural movies contain strong correlations in time (trace under ``natural'' stimulus in Fig.\ \ref{f:stim}) and space (Fig.\ \ref{f:natimages}, \ref{f:stim}).  There are challenges with reliably estimating receptive fields from natural stimuli due to these strong correlations and the highly skewed natural intensity distribution (see Methods).  We therefore also assessed the effect of spatio-temporal correlations in a more controlled stimulus with short-range exponential correlations in time and space and a binary intensity distribution (Fig.\ \ref{f:stim}, ``spat-temp exponential''). 
	
Additional stimuli allowed us to vary the spatial correlation over a broad range, without temporal structure, in order to test the hypothesis that surround strength adapts to remove correlations in nearby parts of an image.  Thus, we examined spatial correlations, in the absence of temporal structure, of increasing extent: spatially exponential, a ``multiscale'' naturalistic stimulus featuring structure over many spatial scales, and full-field flicker (Fig.\,\ref{f:stim}, bottom row).   The multiscale stimulus was designed to mimic the scale invariance of natural scenes in a controlled binary stimulus; featuring both small and large patches of correlated checks (such as the white area near the center).  Its construction is detailed in Methods. In one experiment, we also compared responses to low-contrast white and multiscale stimuli to their high-contrast versions. Finally, to control for the effect of the skewed natural intensity distribution, we also conducted experiments presenting scrambled natural movies lacking spatial or temporal correlation while preserving the intensity distribution.   The mean luminance and single-pixel variance were matched across all stimuli other than natural movies, scrambled natural movies, and low-contrast stimuli. Over 30 minutes of recording in each stimulus condition, the typical cell fired $\sim$7000 spikes. This was sufficient to assess spike train correlations and to measure receptive fields for the white and exponentially correlated stimuli. 

For preliminary analyses, we measured the spike-triggered average (STA) from the ganglion cells' response to white noise.
The resulting receptive fields typically gave good coverage of the sampled visual field (Fig.\ \ref{f:shapes}) and clustered into types on the basis of their response polarity and temporal properties (Fig.\ \ref{f:types}; details in Methods). The four basic types that we consistently identified across experiments were fast-ON and fast-OFF, distinguished by the transient and biphasic nature of their temporal filter, and slow-ON and slow-OFF, which had longer integration times and often less prominent biphasic filter lobes. Separating cells by type did not qualitatively change many of the results reported below; in these cases, we combined all cells to improve statistical power. 

To probe the effect of stimulus correlation on ganglion cell response properties in detail, we applied a standard functional model, the linear-nonlinear (LN) model. In this model, the visual stimulus is filtered with a linear kernel that represents the spatio-temporal receptive field (STRF) of the cell.  The filter output is then passed through a nonlinear transfer function to generate a predicted firing rate.  The nonlinearity encompasses thresholding and saturation, as well as any gain on the linear response.   For white noise stimuli, the STA is a good estimator of the STRF \cite{chichilnisky2001simple}. However, this simple property does not hold for correlated stimuli, and so we fit the STRFs and other LN model parameters by maximum likelihood estimation (see Methods). For the weakly correlated spatio-temporal exponential stimulus, this technique reliably extracted receptive fields (Fig. \,\ref{f:strfs}).

\begin{figure}
\begin{center}
\includegraphics[keepaspectratio, width=\linewidth]{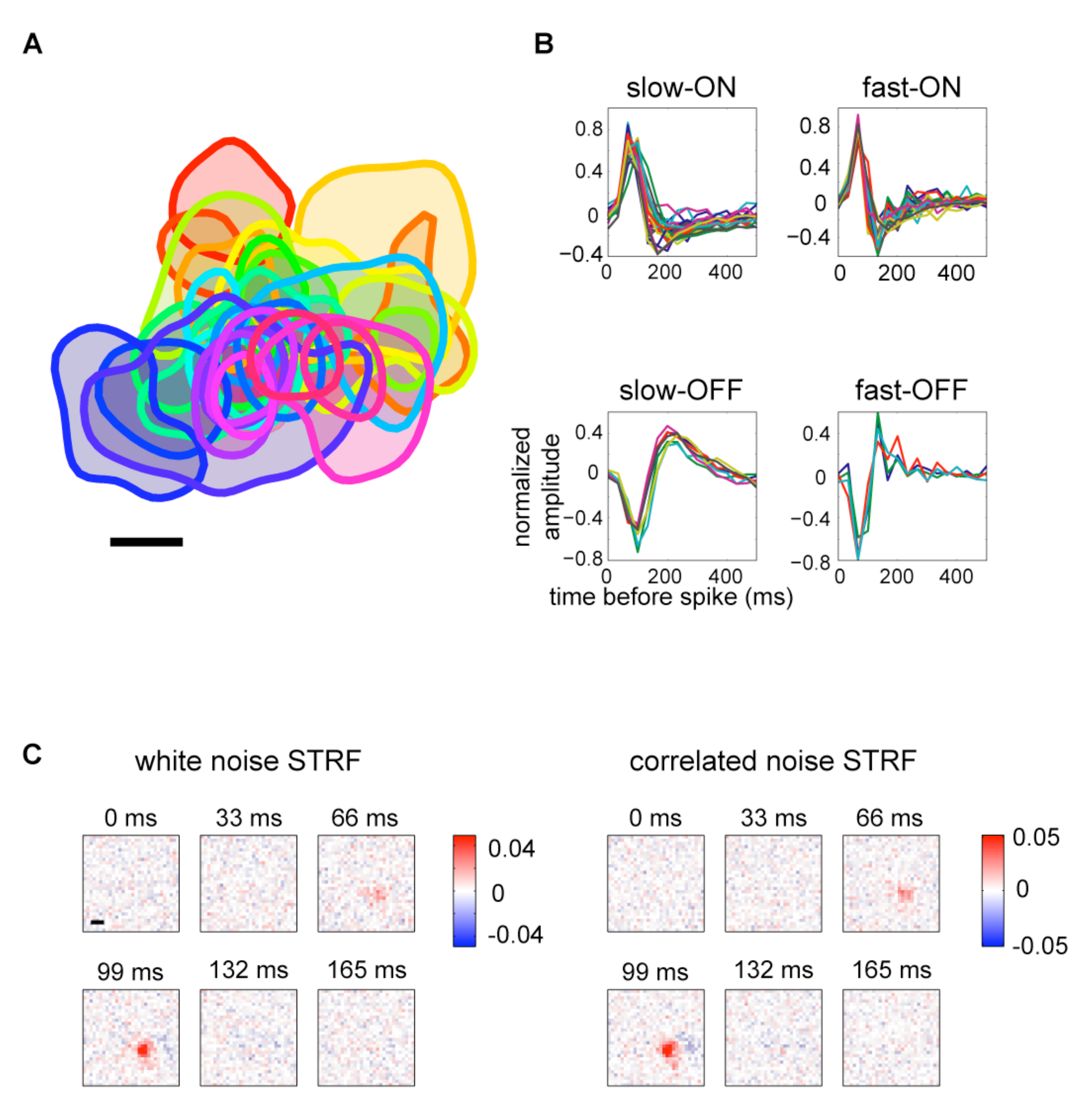}
\phantomsubcaption\label{f:shapes}
\phantomsubcaption\label{f:types}
\phantomsubcaption\label{f:strfs}
\end{center}
\caption{}
\label{Fig2}
\end{figure}

\begin{figure}
\contcaption{
{\bf Retinal ganglion cell receptive fields measured using a multi-electrode array.}  
\textbf{(A)} Receptive field  locations of 31 cells recorded simultaneously from guinea pig retina. Each curve shows the 70\% contour line of one receptive field. Scale bar is $200\ \mathrm{\mu m}$.
\textbf{(B)} Best-fitting temporal kernels for 75 cells, clustered into four types.  Types were obtained by manually clustering temporal filters on the basis of the projection onto their first three principal components.
\textbf{(C)} Maximum likelihood estimates of spatio-temporal receptive fields (STRFs) for an example cell. STRFs were computed separately using responses to white noise (left) or exponential spatio-temporally correlated stimuli (right). Scale bar is $200\ \mathrm{\mu m}$. 
}
\end{figure}

\subsection*{Output correlations are similar between stimulus conditions}
We computed the correlation coefficient between spike trains (binned at 33ms) for all pairs of simultaneously recorded neurons. In response to natural movies, correlations between most pairs of cells increased in magnitude when compared with the correlations between the same pairs when viewing white noise (Fig.\ \ref{f:natcoeff}). We quantified the size of this increase by finding the least-squares best fit line (Fig.\,\ref{f:corrcoeff}, gray lines) and defining the ``excess correlation'' of a population as the slope of this line minus one (see Methods). If all cell pairs had, on average, the same correlation in both stimulus conditions, the excess correlation would be zero. Excess correlation was not strongly dependent on bin size (Fig.\ \ref{f:binsize}). In the case of natural movies, the excess correlation was $0.32\pm 0.20$ (95\% confidence interval computed using bootstrap resampling, as explained in Methods), modestly different from zero. 

\begin{figure}
\begin{center}
\includegraphics[ width=\linewidth]{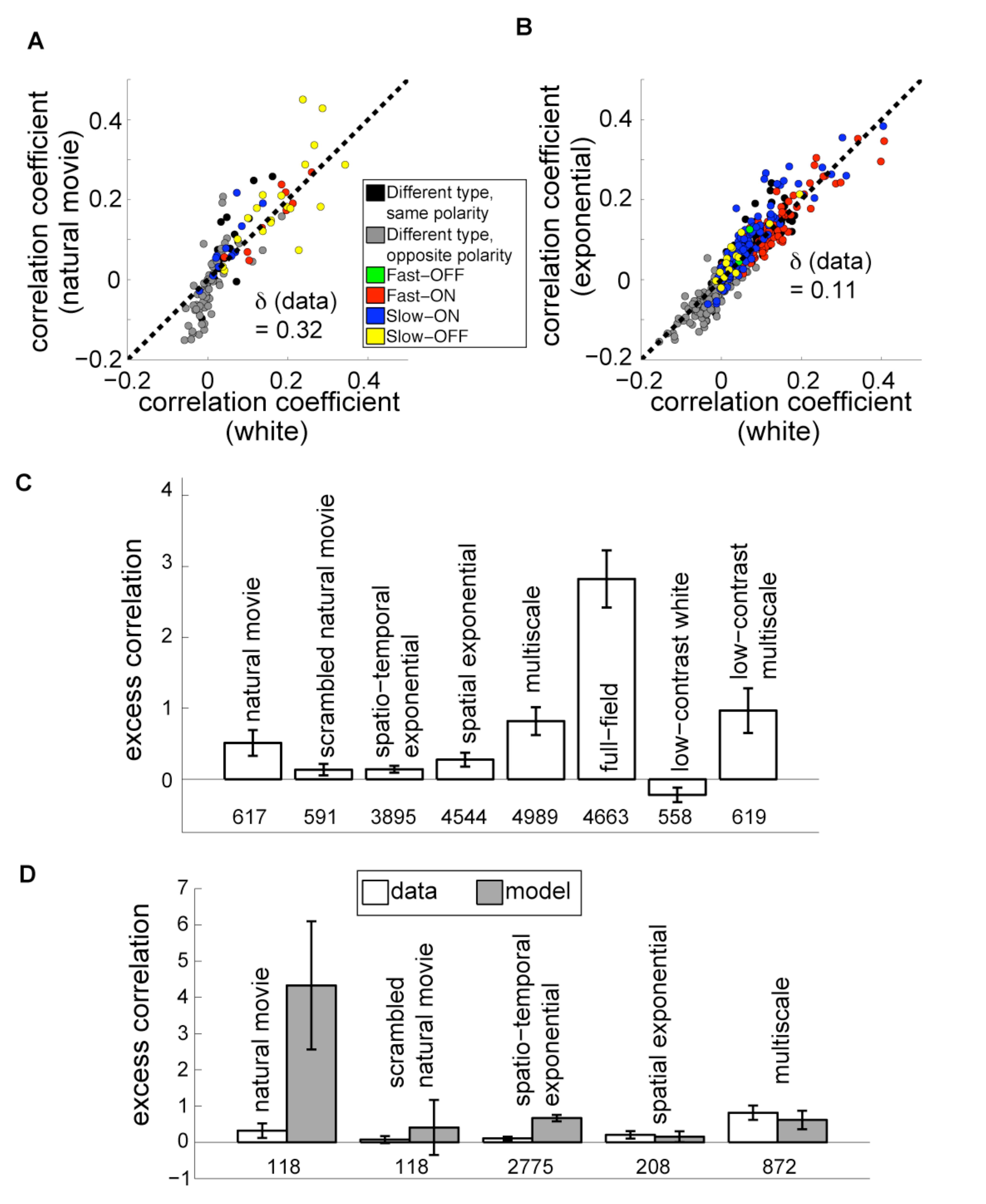}
\phantomsubcaption\label{f:natcoeff}
\phantomsubcaption\label{f:corrcoeff}
\phantomsubcaption\label{f:alldata}
\phantomsubcaption\label{f:allmodel}
\end{center}
\caption{}
\label{Fig3}
\end{figure}

\begin{figure}
\contcaption{
{\bf Retinal output correlations are largely constant between stimulus conditions.}  
\textbf{(A)} Instantaneous spike train correlation coefficients between pairs of ganglion cells, comparing responses to natural movies and to white noise. Dashed black line is the diagonal. Cell pairs of the same type are indicated by colors in the legend.   Different-type pairs  are separated into ON-OFF  (gray) and ON-ON or OFF-OFF pairs (black). The excess correlation, $\delta$, is the deviation of the slope of the best fit line (gray) from the diagonal.
\textbf{(B)} Same as (A) but for spatio-temporal exponentially correlated stimulus. 
\textbf{(C)} Excess correlation measured from ganglion cells responding to the indicated stimulus, compared to white noise.  Numbers below bars indicate the number of cell pairs in each condition; all recorded cells are included. Error bars are 95\% bootstrap confidence intervals computed over 50,000 random samples with replacement from the set of cell pairs. 
\textbf{(D)} Comparison of measured excess correlation (white) to non-adapting model predictions (gray) for the indicated stimuli. Model values were derived from LN neurons with parameters fit to white noise data.  Only cells whose receptive fields met a quality threshold are used here.  
}
\end{figure}

Since the retinal ganglion cell output is a highly transformed representation of its input, it is not trivial to formulate a na\"{i}ve expectation for the magnitude of output correlation given an input correlation.  We therefore chose to quantify this expectation in a simple null model: the LN model fit to the white noise responses.  This model captures correlation due to receptive field overlap and simple nonlinear processing, while neglecting correlations due to shared circuitry and more complex nonlinear behavior, such as adaptation.  For cells which had sufficiently well-estimated white noise LN model parameters (as described in Methods) we were able to compare the excess correlation predicted by the model to that observed in the data. We adjusted the threshold of each model neuron separately under each stimulus to match predicted average firing rates to their empirical values, which differed between stimuli. All other parameters, namely the spatio-temporal receptive field and the gain, were unchanged between stimuli. This ``non-adapting'' model predicted a significantly larger excess correlation in response to natural movies (gray bars in Fig.\ \ref{f:allmodel} and Fig.\ \ref{f:natcorr}), suggesting that the low observed excess correlation value under natural stimulation is a consequence of nontrivial processing in the retina.

In addition to strong correlations, however, natural stimuli are also characterized by a skewed distribution with many dark pixels and a few extremely bright pixels, whereas our white noise stimulus, included equal numbers of bright and dark pixels.  To disentangle effects of correlations from effects due to intensity distribution, we presented the same retinae with a scrambled natural movie.  In this stimulus, we started with natural movies and randomly shuffled the pixels in space and time to maintain the intensity distribution but remove correlations.  The excess correlation in response to this stimulus was consistent with zero in both the measured and simulated responses (Fig.\ \ref{f:natcorr}, left bars).  Moreover, comparing the natural movie and scrambled natural movie directly, we again found a small excess correlation consistent with that in the natural movie vs.\ white noise case.  The non-adapting model again predicts that this low excess correlation is nontrivial (Fig.\ \ref{f:natcorr}, right bars).  Thus, the retina greatly suppresses changes in correlations of natural visual stimuli.

We found a similar set of results for the more weakly correlated spatio-temporal exponential stimulus (Fig.\,\ref{f:corrcoeff}). In particular, the excess correlation was low ($0.12\pm0.05$) whereas simulated responses from the non-adapting model showed an increase (excess correlation of $0.67$; Fig.\,\ref{f:alldata}).  We also examined the results of experiments in which we presented stimuli with varying degrees of spatial correlation (See Table \ref{tab:counts}).  As shown in Figure \ref{f:alldata}, many stimuli produced only a modest increase in output correlations.  Some stimuli with strong spatial correlations, particularly the multiscale and full-field flicker stimuli, resulted in a clear increase in output correlations when compared to white noise.  For stimuli where we varied the contrast (namely white and multiscale noise), output correlations decreased when the contrast was lowered while all other stimulus properties were kept fixed.  Thus, the degree of correlation in the retinal output is not a reflection of stimulus correlations alone.

\begin{table}
\caption{
\bf{Number of cells recorded in each condition.}}
{\small
\begin{center}\begin{tabular}{|l|c|c|c||c|c|c|c|}
\hline
&&&high-quality& \multicolumn{4}{c|}{cell types$^c$}\\
stimulus & retinae & all cells & RFs$^b$ & fast-ON & fast-OFF & slow-ON & slow-OFF\\
\hline
natural movie & 3 & 84 & 34 & 12 & 0 & 9 & 13\\
 & & & & & & &\\
scrambled natural & 3 & 82 & 34 & 12 & 0 & 9 & 13\\
\hspace{4ex}movie & & & & & & &\\
spatio-temporal& 5 & 212 & 75 & 29 & 4 & 31 & 8\\
\hspace{4ex}exponential & & & & & & &\\
spatial exponential$^a$ & 17 & 510 & 46 & - & - & - & - \\
 & & & & & & &\\
multiscale$^a$ & 16 & 513 & 62 & - & - & - & - \\
 & & & & & & &\\
full-field$^a$ & 14 & 483 & - & - & - & - & - \\
 & & & & & & &\\
low-contrast white & 1 & 49 & - & - & - & - & - \\
 & & & & & & &\\
low-contrast & 1 & 49 & - & - & - & - & - \\
\hspace{4ex}multiscale & & & & & & &\\
\hline
\end{tabular}\end{center}
\begin{flushleft} 
$^a$ For our measurements of output correlation (Fig.\ \ref{f:alldata}), we include additional data from experiments performed as part of other studies in which receptive field structure was not probed.  For model correlations and other analyses, we only used the subset of retinae and cells for which we obtained robust receptive field estimates.  \\
$^b$ We used a stringent requirement that receptive fields (RFs) be of high quality for any analyses in which we used receptive field estimates.\\
$^c$ Cells were only divided into subtypes if they had high-quality receptive fields and were recorded in response to stimuli chosen for detailed analysis.
\end{flushleft}}
\label{tab:counts}
\end{table}

\begin{figure}
\begin{center}
\includegraphics[keepaspectratio, width=\linewidth]{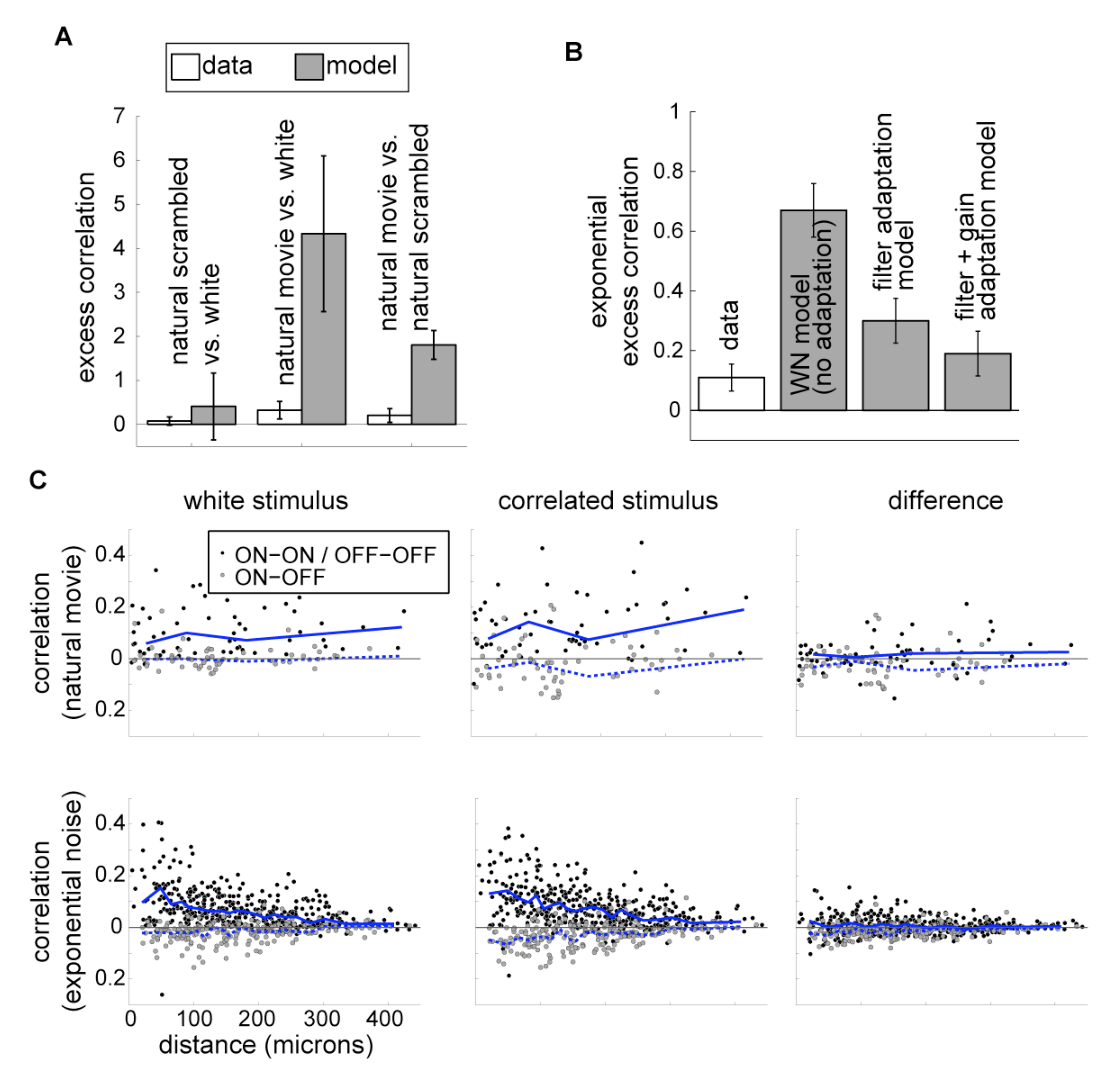}
\phantomsubcaption\label{f:natcorr}
\phantomsubcaption\label{f:adapting}
\phantomsubcaption\label{f:distance}
\end{center}
\caption{}
\label{Fig4}
\end{figure}

\begin{figure}\contcaption{
{\bf Analysis of pairwise correlations.}  
\textbf{(A)}  Excess correlations for natural stimuli. Left and middle bars show excess correlation when scrambled natural movies and intact natural movies, respectively, are compared to white noise in the data and in a population of non-adapting model neurons.  Right bars show excess correlation when responses to natural movies are compared to scrambled natural movies directly.    A non-adapting model predicts larger output correlations in response to the correlated natural input than seen in the data. 
\textbf{(B)} Output correlations under the spatio-temporal exponential stimulus compared with white noise as predicted by LN models with parameters fit to the data. The two leftmost bars (``data'' and ``WN model (no adaptation)'') reproduce the spatio-temporal ``data'' and ``model'' bars in Fig.\ \ref{f:allmodel}.  (Note the difference in scale.) For the other bars, we simulated a population of neurons using linear filters measured from each stimulus but gains measured only from white noise (``filter adaptation model'') or using experimentally derived estimates of both linear filters and gains for each stimulus (``filter + gain adaptation model''). In the fully adapted model, excess correlations are consistent with the data.
\textbf{(C)} Pairwise output correlation as a function of the distance between receptive field centers in the natural movie (top) and spatio-temporal exponential (bottom) datasets. Each dataset contained responses of the same cell population to white noise (left) and to a correlated stimulus (middle). The difference in output correlation in the correlated stimulus over the white noise stimulus is also shown for each cell pair (right).  Each point corresponds to one simultaneously recorded cell pair; the blue lines are the median correlation within bins chosen to contain 30 cells each. Solid lines are median correlations for same-polarity cell pairs; dashed lines are for opposite-polarity pairs. 
}
\end{figure}

For further analysis, we focused on the subset of stimuli shown in Fig.\ \ref{f:alldata}, all of which were presented in experiments where we also obtained robust estimates of white noise receptive fields.  Here we again simulated responses of an LN model using fixed receptive fields measured under white noise.  For all stimuli, the model neurons showed changes in correlation at least as large as those observed in recordings.  However, unlike the spatio-temporally correlated exponential and natural stimuli discussed above, the stimuli which had correlations in space only (spatial exponential and multiscale) or no correlations (scrambled natural movie) produced similar excess correlation values in recorded cells and in our non-adapting model.  This suggests that a fixed linear filter, as in the non-adapting model, is largely sufficient to account for the effect of \emph{spatial} correlations, whereas higher-order processing is necessary to suppress the impact of \emph{temporal} stimulus correlations on output correlation. 

As discussed above, we were able to identify the cell types for many of our recorded neurons.  In response to spatio-temporally exponentially correlated noise and natural movies, cell type had a modest effect on output correlations (Fig.\ \ref{f:natcoeff} and \ref{f:corrcoeff}).  Cells with opposite ON- or OFF- polarities (gray points) tended to have negative correlations, whereas cells of the same polarity (black and colored points) generally had positive correlations.  Moreover, pairs with opposite polarity showed a greater-than-average excess correlation, particularly in response to natural movies. Under natural movies, opposite-polarity pairs had an excess correlation of 1.5; under the spatio-temporal exponential stimulus their excess correlation was 0.38. Within same-type pairs, slow-ON and slow-OFF pairs (blue and yellow) tended to show a greater excess correlation than fast-ON and fast-OFF pairs (red and green). Pairs of slow cells had an excess correlation of 0.29 in the natural stimulus and 0.28 in the spatio-temporal exponential, while fast pairs were measured as 0.01 and -0.02 for the two stimuli, respectively. All of these type-dependent excess correlations are small compared to the overall non-adapting model predictions (excess correlations of 4.33 and 0.67 for natural and spatio-temporal exponential stimuli). We also assessed the relationship between receptive field separation and output correlation (Fig.\,\ref{f:distance}). Pairwise correlations tended to decay with distance, but the change in output correlation between the correlated and white stimulus was small for all receptive field separations. 
	
\subsection*{Adaptation of temporal filters}
	We next sought to determine whether receptive fields adapt to stimulus correlations and whether this adaptation can explain the observed pattern of output correlations.  As noted above, we were able to obtain STAs from responses to white noise.  STAs computed in response to correlated stimuli, however, will be artificially blurred by the stimulus correlations.  To obtain a better estimate of the spatio-temporal receptive field (STRF), we used maximum likelihood estimation to fit a LN model separately for the white and exponentially correlated stimuli \cite{theunissen01}.  Examples of STRFs obtained in this way for one cell are shown in Fig.\ \ref{f:strfs}.  The strongly correlated structure of the multiscale stimulus and the natural movies precluded robust, unbiased STRF estimation with limited data (see Methods).  For this reason, we restricted any STRF computations to white noise and exponentially correlated noise.  The latter stimulus is only weakly correlated and thus we would expect at most weak changes in the receptive fields between the conditions; indeed, receptive fields are hard to distinguish by eye for many cells. Measuring such weak changes requires high-quality receptive fields whose locations can be unambiguously determined (see Methods), as was the case for 75 neurons ($\sim$60\% of the neurons recorded under exponentially correlated conditions). Cells that did not meet this standard were likely to include types that do not respond as well to checkerboard stimuli, e.g., direction selective ganglion cells and uniformity detectors.  We included such cells in the analysis of Fig.\ \ref{f:alldata} in order to maximize the generality of our results and to allow for the possibility that these neurons had qualitatively different output correlations. For the neurons that did pass the quality threshold, we found that the parameters of the LN model (for each neuron, a linear filter and a nonlinearity gain and threshold) changed with the stimulus. 

\begin{figure}
\begin{center}
\includegraphics[width=\linewidth]{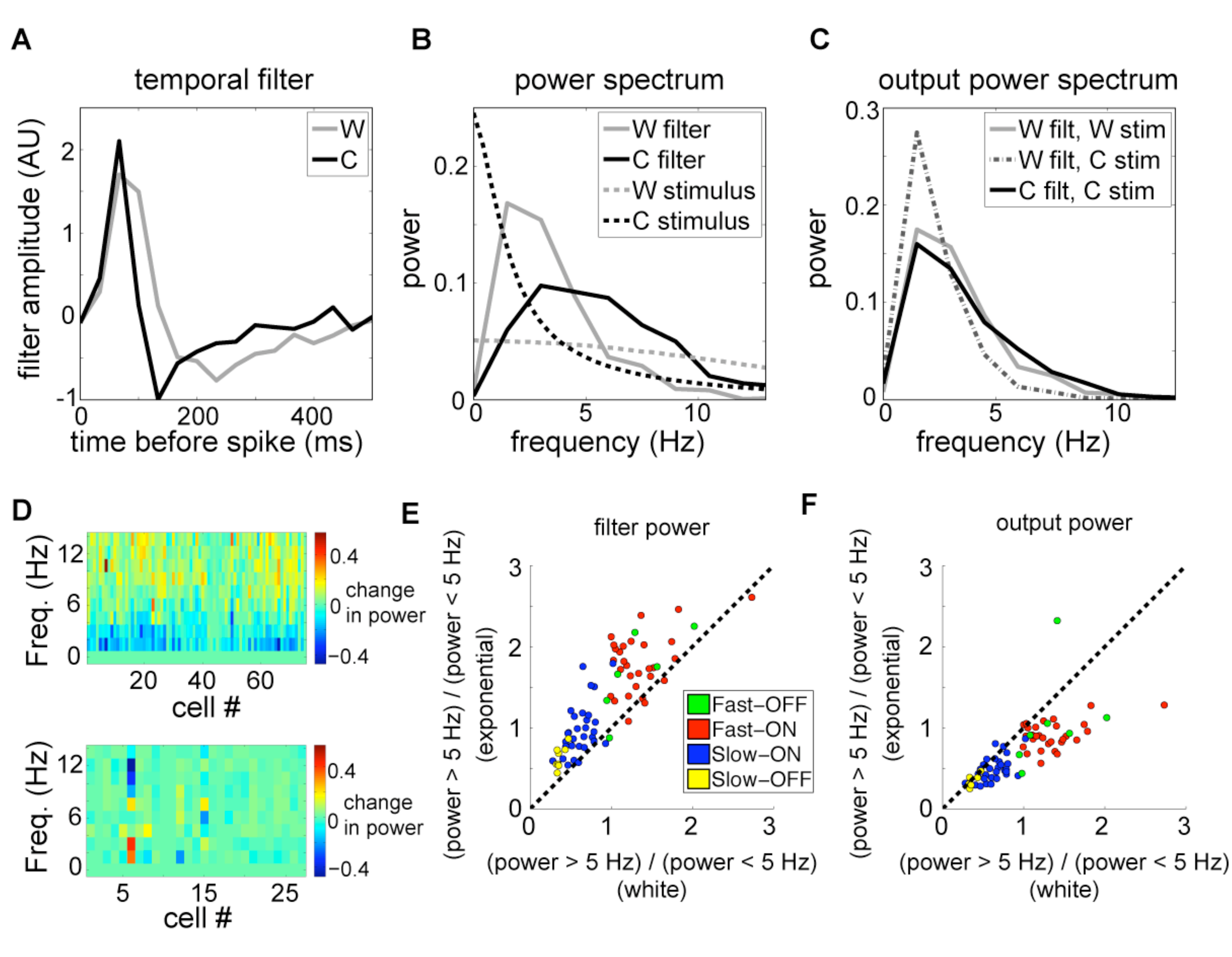}
\phantomsubcaption\label{f:egfilt}
\phantomsubcaption\label{f:eginput}
\phantomsubcaption\label{f:egoutput}
\phantomsubcaption\label{f:allfilts}
\phantomsubcaption\label{f:filtpower}
\phantomsubcaption\label{f:outputpower}
\end{center}
\caption{{\bf Adaptation of the linear temporal filter.}  
\textbf{(A)} Temporal filters are faster under exponentially correlated noise (C) than white noise (W).
\textbf{(B)} Power spectrum of correlated noise input (C, black dashed line) has more low frequency power than white noise (W, gray dashed line).   The power spectrum of the temporal filter for correlated noise (C, black solid line) has more high frequency power. 
\textbf{(C)} Power spectra of filter outputs: White-noise filter acting on  white stimulus (solid gray);  White-noise filter acting on  correlated stimulus (dashed);  Adapting correlated-noise filter acting on  correlated stimulus (solid black). In  adapted cases,  output power spectra are similar between stimuli -- i.e., temporal kernels compensate to maintain invariant output autocorrelation. 
\textbf{(D)} The difference in normalized filter power spectra between the correlated and white stimuli, for spatio-temporal (top) and spatial (bottom) exponential experiments. The power spectra of all filters in each stimulus were normalized by removing the DC component and dividing by the sum of squared amplitudes. The population change in temporal filters shows a consistent increase in high-frequency  power relative to low-frequency power for the spatio-temporal, but not the spatial, stimulus. 
\textbf{(E)} Total power above 5 Hz divided by total power below 5 Hz for filters computed in response to correlated vs.\ white noise stimuli shows a shift towards high-pass signaling across the population. 
\textbf{(F)} Same analysis as in (E) applied to the filter output  in (C). Points near the diagonal indicate near-complete compensation for stimulus changes; points below the diagonal indicate incomplete compensation.
}
\end{figure}

Spike trains with sparse, transient firing events tend to be more decorrelated \cite{pitkow12}. Motivated by this finding, together with our observation that temporally correlated stimuli yielded too-high excess correlation in the non-adapting model, we analyzed adaptation in the temporal filtering properties of retinal ganglion cells. To isolate changes in temporal processing, we examined each neuron's STRFs (estimated separately under the white and exponentially correlated stimulus conditions) and extracted the temporal components (see Methods). These temporal profiles were faster for the correlated stimulus than for white noise (Fig.\ \ref{f:egfilt}). To quantify this difference, we computed the power spectrum of each neuron's temporal filter under each stimulus (Fig.\ \ref{f:eginput} and \ref{f:allfilts}, top) and found a systematic increase in high frequencies under the correlated stimulus, indicating a shift toward high-pass filtering (Fig.\ \ref{f:filtpower}). As the correlated stimulus had relatively more power at low frequencies compared to the white stimulus, this form of adaptation compensates for differences in the power spectrum and, hence, tends to equalize output auto-correlations.  In contrast, a non-adapting model with a filter estimated from white noise acting on the correlated stimulus predicts large changes in the output power spectrum (Fig.\ \ref{f:egoutput}). Indeed, this compensation was nearly exact for many cells (Fig.\ \ref{f:egoutput}), though generally incomplete over the full population (Fig.\ \ref{f:outputpower}).

Next, we found separate temporal profiles for the center and surround and computed the latency, measured as time to peak, of each. Surround latencies did not differ between white noise and spatio-temporally exponentially correlated noise (t-test, $p = .7$, $n = 75$). However, center latencies $l$ were shorter for correlated noise. We quantified the shift in terms of an adaptation index $(l_{corr} - l_{white}) / (l_{corr}+ l_{white})$. The histogram of the adaptation index (Fig.\ \ref{f:latency}; $\mathrm{mean} = -0.03$, $\mathrm{std} = 0.03$; t-test $p <10^{-12}$, $n = 75$; Wilcoxon signed rank text $p<10^{-10}$) showed a robust tail toward shorter center latency for correlated stimuli ($\mathrm{skewness} = -0.53$).  Moreover, almost every cell from which we obtained receptive fields had a longer latency for white noise than for correlated noise (Fig.\ \ref{f:latchange}; $\mathrm{mean\ change} = 6.1\ \mathrm{ms}$).  This was true across cell types.

To determine whether these changes in temporal filtering were due to the presence of temporal correlations in this particular stimulus (unlike many of the other stimuli we examined), we also measured receptive fields from a separate population of ganglion cells responding to white noise and to a stimulus that was exponentially correlated in space but not in time.  In this case, filters did not show a systematic change in power spectra (Fig.\ \ref{f:allfilts}, bottom), but the center latencies were still shorter for the correlated stimulus (Fig.\ \ref{f:latchange}; $\mathrm{mean\ change} = 7.2\ \mathrm{ ms}$).  Again, computing adaptation indices indicated that this effect was significant ($\mathrm{mean} = -0.04$, $\mathrm{std} = 0.03$; t-test $p <10^{-10}$, $n = 37$; Wilcoxon signed rank test $p<10^{-7}$).  Thus, spatial correlations in the stimulus affect the temporal filtering properties of neurons, albeit to a less degree than spatio-temporal correlations.  These results, combined with those of \cite{pitkow12}, may indicate that the observed consistency of correlations is produced by an increase in response transience when stimulus correlations increase.

\subsection*{Adaptation of spatial receptive fields and nonlinearity gain}
The conventional view of retinal circuitry suggests that adaptive decorrelation arises from stronger or wider surround inhibition during viewing of correlated stimuli. We thus computed the amplitudes of the surround and center components of each neuron's STRFs in both white noise and spatio-temporally exponentially correlated noise. Defining the relative surround strength, $k$, as the ratio of surround amplitude to center amplitude (details in Methods), we computed an adaptation index for each cell as $(k_{corr} - k_{white}) / (k_{corr} + k_{white})$. This adaptation index has a modestly positive mean (Fig.\ \ref{f:surround}; $\mathrm{mean} = 0.075$, $\mathrm{std} = 0.24$; two-tailed t-test, $p = 0.008$, $n = 75$; Wilcoxon signed rank test $p=.008$), as do the changes in surround strength themselves (Fig.\ \ref{f:surrchange}).  In addition, the cells with the greatest degree of surround adaptation had a robust tendency to increase in surround strength ($\mathrm{skewness} = 0.15$).  There was no discernible dependence on cell type.  Interestingly, the surround strength showed only a marginally significant change when spatial correlations (but not temporal correlations) were added to white noise (Fig.\,\ref{f:surrchangespatial}; $\mathrm{mean\ adaptation\ index} = -0.087$, $\mathrm{std} = 0.26$; two-tailed t-test, $p = 0.05$, $n = 37$; Wilcoxon signed rank test $p=.02$). Thus, while we do find some evidence for an increase in surround strength with stimulus correlation, the effect is subtle. This outcome is surprising given the prevailing view since the work of Barlow \cite{barlow61, srinivasan82} that surround inhibition is primarily responsible for decorrelation of visual stimuli. However, it is possible that the exponential stimulus that permitted us to estimate receptive fields is too weakly correlated to evoke strong lateral inhibition.

\begin{figure}
\begin{center}
\includegraphics[width=\linewidth]{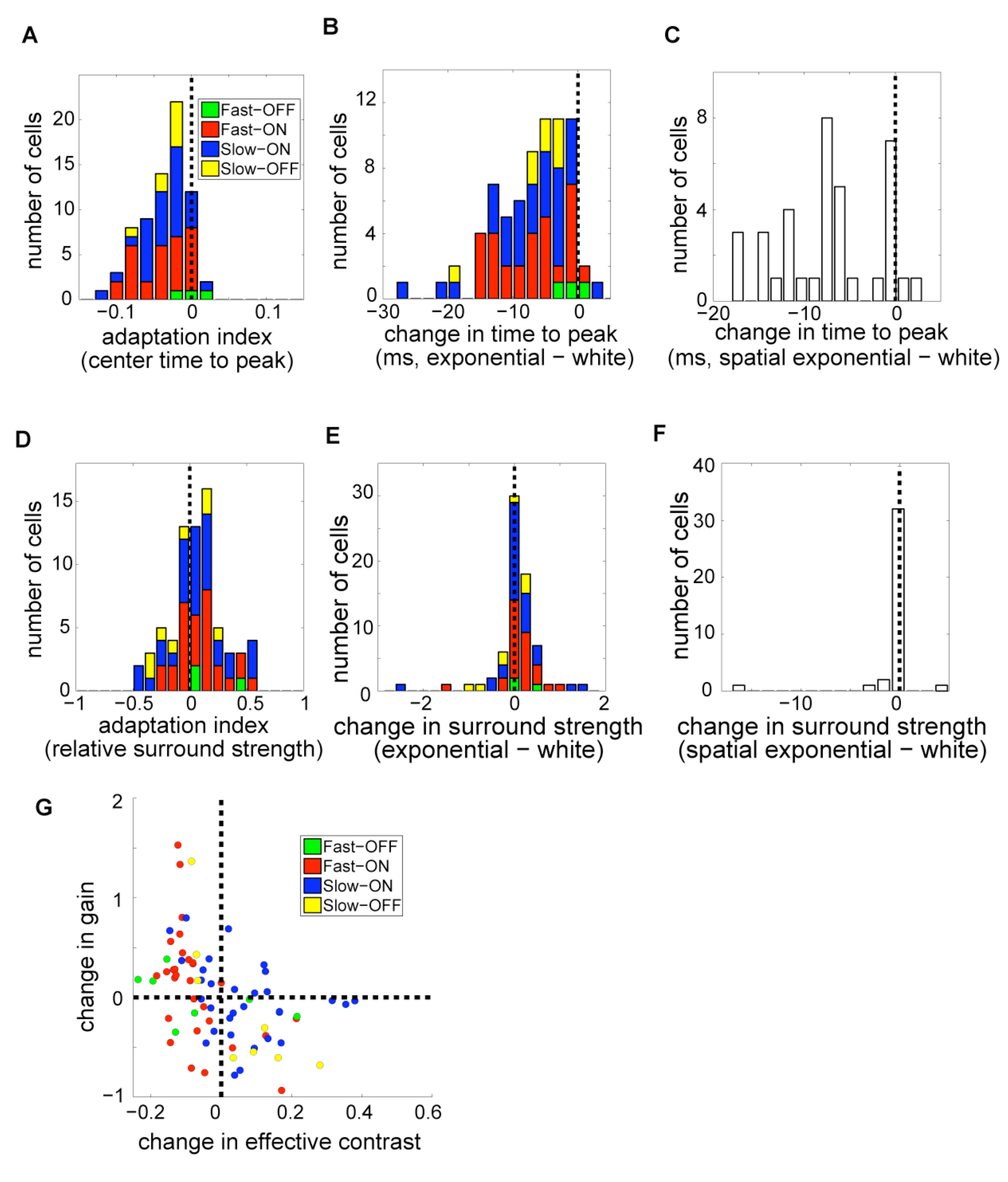}
\phantomsubcaption\label{f:latency}
\phantomsubcaption\label{f:latchange}
\phantomsubcaption\label{f:latchangespatial}
\phantomsubcaption\label{f:surround}
\phantomsubcaption\label{f:surrchange}
\phantomsubcaption\label{f:surrchangespatial}
\phantomsubcaption\label{f:gainchange}
\end{center}
\caption{}
\label{Fig6}
\end{figure}

\begin{figure}\contcaption{
{\bf Adaptation of the spatio-temporal receptive field and gain.}
\textbf{(A)} Center latency (time to peak of the temporal kernel) is shorter for exponentially correlated noise. Histogram shows adaptation indices (corr-white) / (corr+white) for center latency ($n = 75$). 
\textbf{(B,C)} Changes in center latency (corr-white) for spatio-temporally correlated (B) and temporally correlated (C) stimuli, in milliseconds.  Almost all cells have a decreased time to peak when responding to a correlated stimulus. 
\textbf{(D)} Adaptation indices, computed as in (A), for relative surround strength (surround/center ratio) show a slight skew toward a stronger surround for spatio-temporally  correlated noise. ($n = 75$). 
\textbf{(E, F)} Difference in surround strength for the spatio-temporal (E) and spatial (F) exponential stimuli.
\textbf{(G)} Gain adaptation. Gains were defined as the slope of the LN model nonlinearity, and obtained separately for the response to white noise and to the spatio-temporally correlated exponential stimulus. Effective contrast, the standard deviation of the linear filter output, was similarly measured in both stimuli. The difference in gain, correlated value minus white noise value, is plotted against the difference in effective contrast. Increases in effective contrast tend to invoke compensating decreases in gain. ($n=75$). 
}
\end{figure}

Finally, we examined the gain $g$ of each neuron, defined as the maximum slope of the logistic nonlinearity fit to each neurons' response (see Methods). Since the gain enters the nonlinearity after the stimulus passes through the linear filter, we normalized the filter to unit euclidean norm in order to obtain an unambiguous definition of $g$.  We found that the gains of individual neurons changed when the stimulus was more correlated, but there was not a systematic change between stimuli.  Recall that the gain of many sensory neurons, including retinal ganglion cells, is known to change with the contrast of the stimulus \cite{smirnakis97, baccus02}. To test for  a possibly related mechanism at work in our data we first defined ``effective contrast,'' $\sigma_{white}$ and $\sigma_{corr}$, as the standard deviation of the normalized linear filter output in each stimulus, respectively.  This notion of effective contrast roughly captures the variability of the ganglion cells' input, taking presynaptic processing into account. Any nonlinear gain control mechanism in the ganglion cell layer should therefore be sensitive to this quantity. For some cells $\sigma_{corr}$ exceeded $\sigma_{white}$, while for others the reverse was true. Measuring the gains in both stimulus conditions ($g_{white}$ and $g_{corr}$), however, we found systematic adaptation opposing the change in effective contrast: gain tended to increase when effective contrast decreased and vice-versa.  Specifically, the quantities $\Delta g = g_{corr} - g_{white}$ and $\Delta\sigma = \sigma_{corr} - \sigma_{white}$ were significantly anticorrelated (Fig.\, \ref{f:gainchange}; Spearman's $\rho = -0.54$, $p = 10^{-6}$, $n = 75$).

\subsection*{Output correlations in an adapting model}
Finally, we assessed whether the receptive field changes reported above could account for the observed similarity of output correlations between white noise and the spatio-temporal exponential stimulus.  For experiments using spatio-temporally exponential noise, as discussed above, we measured the adaptation in LN model parameters fit to each stimulus. We then separately examined the effect of adaptation in different parameters on the excess correlations predicted by the LN models. Including adaptation of the linear filters, but not the gain, produced a significantly improved match between the model and the data (Fig.\ \ref{f:adapting}, ``filter adaptation model''). Additionally allowing the gain to adapt produced output correlations consistent with the data (Fig.\ \ref{f:adapting}, ``filter + gain adaptation model'').  The contribution of gain adaptation to decorrelation is interesting in light of our observation that output correlations are lower for stimuli with lower contrast (Fig.\ \ref{f:alldata}).  Low contrast stimuli generally evoke lower firing rates, which could result in decreased pairwise correlations absent any change in linear filtering properties. (See Text \ref{para:s1} for a derivation of this result.)  At the same time, changes in contrast lead to gain control, wherein gain is higher for lower stimulus contrast.  This gain adaptation could also affect output correlations, as in Fig.\ \ref{f:adapting}.  It would be interesting to know how gain control interacts with changes in other properties, such as the nonlinearity threshold and the shape of the linear filter, to set the correlations in the retinal response.  Note that the LN model is fit to each neuron independently, without taking correlations between neurons into account. Its successful prediction of the change in pairwise correlations, without explicit introduction of inter-neural interactions, is therefore noteworthy.  We conclude that observed adaptation in receptive fields and gains is adequate to explain the output correlations in responses to a spatio-temporally correlated stimulus.

\section*{Discussion}

Our principal finding is that the retina maintains a moderate, and relatively constant, level of output correlation across a wide range of  spatio-temporally correlated stimuli ranging from white noise to  natural movies.   Our data also suggest a differential effect of spatial vs. temporal correlations on the functional properties of the retinal output.      We focused here on spatial variations in our control stimuli, but it would interesting to design future studies to explore the space-time differences more systematically.  In response to exponentially correlated noise, where the receptive fields could be estimated, we showed that the relative invariance of output correlations is largely accounted for by the observed changes in the linear receptive field (faster temporal kernels and slightly stronger surround inhibition for more correlated stimuli) and by changes in the nonlinear gain (anti-correlated to changes in effective contrast). While the latter findings give an interpretation of the results in terms of a conventional functional model (here a linear-nonlinear cascade), the measurement of output correlations is model-independent.

Classifying cells into types revealed a slight dependence of excess correlation on cell type: most robustly, opposite polarity ON-OFF pairs showed the greatest increase in correlation magnitude when stimulus correlation increased. Indeed, if the retinal output is split across parallel functional channels, redundancy is likely to be highest within a channel due to shared circuit inputs.  It may thus be advantageous, from an information encoding perspective, for decorrelation to act \emph{within} a channel, with residual correlations across types signaling to downstream areas relevant relationships between the information in different channels.

Pitkow and Meister \cite{pitkow12} showed that retina partly decorrelates naturalistic inputs but that the response to white noise is more correlated than the input, in part due to receptive field overlap between ganglion cells.  Consistent with their results, we found that changes in output correlations were often smaller than changes in input correlations.  We also extended their findings by showing that this partial decorrelation occurs in individual pairs of neurons. In \cite{pitkow12}, it was also suggested that the linear receptive field measured from white noise was insufficient to explain the amount of decorrelation seen for naturalistic stimuli and the bulk of the decorrelation was attributed to changes in the threshold of a functional model of ganglion cells. However, the authors did not directly measure the (possibly different) receptive fields of ganglion cells responding to correlated stimuli, nor did they follow particular cell pairs across different stimuli. Our measurements suggest that the nonlinear processing proposed in \cite{pitkow12} can be described in terms of adaptation of the linear receptive field and nonlinear gain with the net effect that output correlations are roughly constant for each cell pair across a range of correlated stimuli, as was observed in visual cortex by \cite{fiser04}. Our results also recall those of \cite{lesica07}, \cite{david04}, \cite{sharpee06}, and \cite{touryan05}, who showed that receptive fields in LGN and primary visual cortex differ in structure when probed with natural movies versus random stimuli. 

We also found that the gain of retinal ganglion cells responding to correlated stimuli changes with ``effective contrast'' $\sigma_{white}$ and $\sigma_{corr}$, i.e.\ with the standard deviation of the input to the nonlinearity in a linear-nonlinear model of ganglion cells. In classical contrast gain control, firing rates and response kinetics adapt to temporal contrast and to the spatial scale of stimuli \cite{smirnakis97, baccus02}.  As increased stimulus correlation may produce a qualitatively similar input to the inner plexiform layer as increased contrast, some of the cellular mechanisms underlying contrast adaptation might also contribute to the phenomena we have uncovered. This provides an avenue for future study of the functional mechanisms underlying adaptation to correlations.

We have focused in the present work on the failure of a non-adapting linear-nonlinear model to capture the small scale of observed excess correlations and have seen that adaptation in the linear filter might remedy this discrepancy.  Alternatively, shared circuitry in the population of neurons may be engaged by correlated inputs and require explicit inclusion in any functional model of retinal responses to different classes of correlated stimuli \cite{ganmor11, pillow08}.    Such shared circuitry leads to noise in one neuron being passed to multiple nearby neurons, and is thus measured by ``noise correlations.''  While addition of fixed, stimulus-independent noise correlation would not greatly change our results, a \emph{change} in noise correlation with stimulus would provide a different candidate mechanism to account for our data \cite{granot-atedgi13}.  This is another avenue for future work. 

We have focused here on the effects of spatial correlations in an experimental design where we could compare receptive fields computed from responses to two different stimuli.  Thus, we used relatively weak exponential correlations to ensure that we were not measuring artifacts of the stimulus correlations themselves. Recovering receptive fields from strongly correlated stimuli can require long recording times.  Thus, because our experimental design involved comparisons between several different stimuli, we were only able to recover receptive fields for moderately correlated stimuli.  Future work could simply present each stimulus for a longer duration to assess receptive field changes at a population level rather than analyzing multiple stimuli in one experiment.  Further work could also include parallel studies with stimuli including temporal correlations only to complement our findings on responses to spatial correlations.  

Finally, it would be interesting to determine the timecourse of the adaptations observed here.  Knowing whether a change in stimulus correlations induces changes in receptive fields and output correlations within seconds, tens of seconds, or longer would help to clarify the relationship between processing of correlations and adaptation to other stimulus features such as contrast.  Again, the design of our experiments precluded making these measurements -- we focused on long segments to measure steady-state processing of correlations, whereas assessing the timecourse of changes requires finer and more systematic sampling of transitions between stimuli.

Why would the retina need to adapt, in the behaving animal, to variations in spatial correlations? While natural scenes are scale-invariant on average, the specific correlations do vary depending on the scene and the viewing distance (see Fig.\ \ref{f:natimages}). Barlow originally suggested that sensory systems should decorrelate their inputs to make efficient use of limited neural bandwidth \cite{barlow61}. Consistent with this idea, we found that retina removes redundancies, in spatio-temporally correlated stimuli, but also that the retinal output is not completely decorrelated. Rather, the output correlations are reduced to an approximately fixed level, roughly matching  correlations in responses to white noise checkerboards.   What drives this tradeoff? Recall that redundancy can be useful to protect against noise, to facilitate downstream computations, or to enable separate modulation of information being routed to distinct cortical targets. Thus, it may be that a certain degree of output correlation between retinal ganglion cells represents a good balance between the benefits of decorrelation and the benefits of redundancy \cite{tkacik10}. Sensory outputs with varying amounts of correlation may also be decoded differently by cortex \cite{estebanez12}, in which case maintaining a fixed visual code might require that retinal output correlations are within the range expected by downstream visual areas. In these interpretations, it makes sense that the retina adapts to maintain correlation within a narrow range across stimulus conditions, as we have found.

\section*{Methods}

\paragraph{Ethics statement.} All procedures were in keeping with the guidelines of the University of Pennsylvania, the NIH, and the AVMA.

\paragraph{Neural recording.} We recorded retinal ganglion cells from Hartley guinea pig using a 30-electrode array ($30\ \mathrm{\mu m}$ spacing, Multi Channel Systems MCS GmbH, Reutlingen, Germany). After anesthesia with ketamine/xylazine (100/20 mg/kg) and pentobarbital (100 mg/kg), the eye was enucleated and the animal was euthanized by pentobarbital overdose. The eye was hemisected and dark adapted. The retina was separated from the pigment epithelium, mounted on filter paper, and placed onto the electrode array, ganglion cells closest to the electrodes. Extracellular signals were recorded at 10 kHz.  Spike times were extracted with the spike-sorting algorithm described in \cite{prentice11}; briefly, a subset of data was manually clustered to generate spike templates that were then fit to the remaining data using a Bayesian goodness-of-fit criterion.

\paragraph{Stimulus generation.} 
We displayed checkerboard stimuli (see Fig.\ \ref{f:stim}) at 30 Hz on a Lucivid monitor (MicroBrightField inc., Colchester, VT) and projected the image onto the retina. The mean luminance on the retina was $9000\ \mathrm{photons/s\cdot\mu m}^2$ (low photopic); each check occupied an area between $50\ \mathrm{\mu m}$ x $67\ \mathrm{\mu m}$ and $100\ \mathrm{\mu m}$ x $133\ \mathrm{\mu m}$.  To make white noise and exponentially correlated stimuli, we first produced random checkerboards with intensities drawn from a Gaussian distribution. Spatio-temporally correlated stimuli were produced by filtering the Gaussian stimulus with an exponential filter with a time constant of three stimulus frames (99 ms) and a space constant of one check to match the scale of typical receptive fields.  Stimuli with only spatial exponential correlations were constructed similarly, but with a time constant set to zero. To create the multiscale stimulus, we first generated gaussian white noise checkerboards at several power-of-two scales.  The largest scale consisted of a single check filling the entire stimulus field, the next largest was a 2 x 2 checkerboard (with check size equal to half the stimulus field), the third largest was a 4 x 4 checkerboard (check size one quarter of the stimulus field), and so on. The checkerboards at all scales were then summed and thresholded to obtain a binary stimulus qualitatively mimicking the scale-invariant structure of  spatial correlations in natural scenes (Fig. \, \ref{f:stim}). This stimulus did not contain temporal correlations.  Natural movies of leaves and grasses blowing in the wind were collected with a Prosilica GE 1050 high-speed digital camera with a 1/2'' sensor (Allied Vision Technologies GmbH, Stadtroda, Germany) connected to a laptop running StreamPix software (NorPix Inc., Montreal, Canada) to grab frames at 60 fps.  The camera resolution was set to 512 x 512 pixels, and movies were filmed from a fixed tripod about 5 feet from the trees and grass.  Natural light was used to illuminate our outdoor scenes, and exposure time was set ($300-3000\ \mathrm{\mu s}$) to capture variation in shadows and avoid saturation of our 8-bit luminance depth.  Videos were collected for up to 30 minutes; 10 second to 5 minute segments with continuous motion were selected.  Videos were downsampled to match the resolution and frame rate of our stimulus monitor.  To produce a scrambled control for natural movies, pixels were randomly shuffled in space and time to remove all correlations.  All stimuli other than natural movies (intact and scrambled) were thresholded at the median to fix the mean luminance and single-pixel variance and to maximize contrast. This binarization did not affect the power spectra significantly. For low-contrast stimuli, all deviations from the mean luminance were halved to give an overall contrast of 50\%. Typically, we alternated 10-minute blocks of white noise with 10-minute blocks of a correlated stimulus.

\paragraph{Cell typing.} We used reverse correlation to compute the spike-triggered average (STA) for each cell responding to white noise. We performed principal component analysis (PCA) on the best-fitting temporal kernels and split cells into 
two clusters based on the sign of the first component; the clusters were identified as ON and OFF types based on the sign of their temporal kernels. (Our under-sampling of OFF cells \cite{borghuis08, ratliff10} may be due to electrode bias, as individual OFF cells are smaller and therefore less likely to be detected by our electrode array.) PCA was repeated for the ON and OFF groups separately. We manually identified clusters based on the projections onto the first three principal components; in this way we identified four functional types, including  slow-OFF, fast-OFF,  fast-ON, and  slow-ON (see Fig.\ \ref{f:types}).

\paragraph{Maximum likelihood estimation of linear-nonlinear models.} 
To obtain spatio-temporal receptive fields (STRFs) for both white and exponentially correlated stimuli, we used publicly available code (strflab.berkeley.edu; \cite{theunissen01}) to carry out maximum likelihood estimation. We parameterized the model by a linear filter acting on the stimulus and a logistic nonlinearity, so that firing probability is $p(s) = 1 / (1 + \exp(-g*(s-b)))$, where $s$ represents the linear filter output, and $g$ and $b$ are gain and offset parameters. Gradient ascent with early stopping was used to compute a maximum likelihood estimate of the linear filter that best fit the data.  We initialized the algorithm for each neuron using the spike-triggered average recorded in response to white noise.  Many cells do not yield clear receptive fields when probed with white noise, either because this stimulus does not evoke a sufficiently strong response or because the response is not well modeled as a single linear filter.  To avoid potential artifacts that could arise from including such cells in our receptive field and model analyses, we selected cells whose receptive fields had clearly visible centers.  This classification of receptive fields as high-quality was done before any other data analysis in order to avoid biasing the selection.  In datasets where we obtained receptive fields for both white noise and a correlated stimulus the designations of high-quality agreed between the two stimuli for 98\% of cells.  The subset of cells identified in this way also had center locations that were clearly delineated by our automated receptive field analysis, giving confirmation of our visual threshold.

The STRF baseline was poorly constrained by the maximum likelihood procedure, since an additive change in the STRF has a similar effect to a proportional shift in the offset parameter of the nonlinearity.  We therefore normalized the STRFs by subtracting an estimate of the baseline: we allowed the fit to include components extending 100 ms after the spike --- where the true filter must be zero by causality --- and subtracted the mean of these frames.  Inclusion of these post-spike frames also allowed us to verify that the temporal autocorrelations in our stimuli did not produce any acausal artifacts in the recovered STRFs. We normalized the estimated linear filters to have unit Euclidean norm (square root of the sum of squares of filter values) and then used gradient ascent to separately fit the gain and offset of a logistic nonlinearity.  Since the likelihood function in this case is convex, there was no possibility of local maxima.  While we were able to compute unbiased estimates of STRFs from responses to stimuli with exponential correlations, our multiscale and natural movie stimuli were too correlated to estimate unbiased receptive fields with the number of spikes we were able to obtain in a single recording.

\paragraph{Correlation analysis.}  Correlations were measured as the correlation coefficient between pairs of simultaneously recorded neurons.  Spike trains were divided into 33 ms bins; we assigned a bin a 1 if it had one or more spikes and a zero otherwise.  The results reported above did not change if we used spike counts in each bin rather than binarizing. Indeed, 98\% of timebins had one or fewer spikes and less than 0.05\% had more than three spikes. We summarized the results by finding the best fit line of the form $\rho_{corr} = (1+\delta) \rho_{white}$, where $\rho_{white}$ and $\rho_{corr}$ are the pairwise correlations under the white and correlated stimuli, respectively. We estimated the excess correlation, $\delta$, by the total least squares regression method and computed 95\% bootstrap confidence intervals from 50,000 bootstrap resamples of the set of ganglion cell pairs.

Such instantaneous correlations are thought to combine slow stimulus-driven effects with fast intrinsic effects due to shared noise \cite{greschner11}.  To verify that this did not affect our results, we isolated the stimulus-driven component, by noting that our cross-correlation functions can feature a short-timescale peak riding on a slow component and extracting the latter.  Specifically, we binned the spike trains into 1 ms bins and computed cross-covariance functions between pairs.  To isolate the stimulus-induced component, we smoothed the cross-covariance functions by fitting a cubic B-spline curve with knots spaced at 20 ms to suppress the fast noise component.  We then found the shift, within a 200 ms window, which maximized the absolute value of the smoothed cross-covariance and estimated the correlation coefficient as the cross-covariance at this shift normalized by the product of the standard deviations.  This gave excess correlation values consistent with those reported above (not shown).

We also computed the power spectra of the stimuli, the best-fitting temporal kernels, and the filter outputs (i.e.\ stimulus power spectra multiplied by filter power spectra). We summarized each power spectrum by computing the total power above 5 Hz divided by the total power below 5 Hz.

\paragraph{Measures of receptive field characteristics.} Given a STRF estimated for one cell under one of the stimulus conditions, we first performed principal component analysis on the timecourses of the individual pixels.  From the resulting set of ``principal timecourses'' we selected the one most similar to the timecourse of the pixel that achieves the peak value in the full STRF.  The complete linear filter was collapsed into a single frame by finding the projection of each pixel onto this principal timecourse. This procedure is equivalent to finding the best (least squares) spatio-temporally separable approximation to the STRF: $K(x,t) = k(x) w(t)$, where $k(x)$ and $w(t)$ are the spatial and temporal components of the approximation.  From the single frame $k(x)$, we extracted the center and surround regions.  To find the center, we began with the peak pixel and then recursively expanded the region in a contiguous patch to include any pixels whose timecourses had at least a 50\% correlation with already included pixels.  We ended the recursive process after the first pass in which no pixels were added to the center.  At this point, all pixels not included in the center were considered part of the surround for the purpose of assessing the surround strength.  

Taking the center defined in this way as a mask for the full STRFs, we summed all pixel values within the center at each time point to generate a temporal profile for the central receptive field. To obtain temporal kernels with greater precision than the 30 Hz time scale of our STRFs, we used cubic spline interpolation with knots spaced every 33 ms. From the interpolated timecourses, we measured the time to peak under each stimulus for the center. In addition, the peak value of this temporal profile was taken to be the center weight of the receptive field. Similar computations yielded the surround time to peak and surround weight. We then quantified the relative surround strength as the ratio of surround weight to center weight. 

In addition, we measured the gain $g$ of each neuron under each stimulus condition. To test for contrast gain control, we defined ``effective contrast,'' $\sigma$, as the standard deviation of the linear filter output.  To avoid ambiguity between linear filter amplitude and gain, we normalized each STRF to have unit Euclidean norm before computing the gain and the effective contrast.
	
We used the analysis method described here because it gave the most robustly unbiased results in our simulations (see below), but we also wanted to verify that our results did not change dramatically with slightly different methods (see details in Text \ref{para:s2} and Table \ref{tab:data}).  
Briefly, we made a series of modifications to our receptive field extraction method and repeated the analyses described in the main text for each modification.  

\paragraph{Model validation.}  
\begin{table}
\caption{
\bf{Model validation of receptive field analysis.}}
{\small
\begin{center}\begin{tabular}{|l|c|c|c|c|c|c|c|c|}
\hline
& \multicolumn{4}{c|}{AI (relative surround strength)} & \multicolumn{4}{c|}{AI (center time to peak)}\\
& mean & std & $p$ & skew & mean & std & $p$ & skew\\
\hline
Standard model & $-0.02$ & 0.06 & .004 & $-1.10$ & $-0.0003$ & 0.002 & .04 & $-4.70$\\
Small surround weight & $-0.06$ & 0.13 & $<.0001$ & $-0.12$ & $-0.0003$ & 0.002 & .04 & $-4.69$\\
Large surround weight & $-0.05$ & 0.26 & .08 & $-2.44$ & $-0.0014$ & 0.004 & .0003 & $-2.66$\\
Small surround radius & $-0.01$ & 0.08 & .19 & $-4.56$ & $-0.0009$ & 0.003 & .0007 & $-2.49$\\
Large surround radius & $-0.02$ & 0.05 & $<.0001$ & $-0.24$ & $-0.0005$ & 0.002 & .01 & $-3.71$\\
\hline
\end{tabular}\end{center}
\begin{flushleft}
Adaptation index (AI) in surround strength and center latency for different non-adapting control models. Columns labeled ``mean,'' ``std,'' and ``skew'' show the mean, standard deviation, and skewness of the adaptation indices for the indicated analysis; columns labeled ``$p$'' show the $p$-values from $t$-tests of each distribution against the null hypothesis of zero mean. \textbf{Standard model}: Surround radii (relative to center radii) had mean 2 and standard deviation 0.3; surround weights (relative to center weights) had mean 1 and standard deviation 0.1. \textbf{Small surround weight}: Surround weights had mean 0.5; all other parameters were the same as in the standard model. \textbf{Large surround weight}: Surround weights had mean 2. \textbf{Small surround radius}: Surround radii had mean 1. \textbf{Large surround radius}: Surround radii had mean 3.
\end{flushleft}}
\label{tab:model}
 \end{table}

To validate our STRF analysis methods, we generated synthetic data using a linear-nonlinear (LN) model.  We then applied STRF extraction and analysis methods identical to those applied to real data to check that the known LN parameters were extracted in an unbiased fashion.  The linear filter was chosen to be spatio-temporally separable, with the temporal component taken from measured ganglion cell responses. The spatial filter was modeled as a difference-of-Gaussians, where the size and strength of the surround Gaussian relative to the center Gaussian were chosen to mimic receptive fields of real neurons.  In each simulation, parameters for 100 model neurons were chosen independently.  The results are summarized in Table \ref{tab:model}.

In our first simulation, the surround radius (relative to center radius) was chosen from a Gaussian distribution with mean 2 and standard deviation 0.3, the relative surround strength from a Gaussian distribution with mean 1 and standard deviation 0.1, and the offset coordinates from Gaussian distributions with mean 0 and standard deviation 2 (``Standard model'' in Table \ref{tab:model}).  For each model neuron, the same filter was applied to the exponentially correlated and uncorrelated stimuli in order to simulate cases without adaptation. Across the population, our model neurons showed only a slight bias in center latency between the two stimuli (Fig.\ \ref{f:latmodel}).  While this effect reaches significance (for $\alpha = .05$), the effect size is orders of magnitude smaller than that seen in the data and thus could not explain our experimental results.  We also observed a tendency toward a slightly stronger relative surround strength under white noise than under correlated noise (Fig.\ \ref{f:surrmodel}).  Note that this is opposite the effect observed in our experimental results (Fig.\ \ref{f:surround} -- \ref{f:surrchangespatial}).  Thus, if anything our results may be stronger than reported in the main text.  

To further validate our analysis we ran simulations with an even wider range of model parameters.  We first constructed model neurons with surround radii measured from Gaussian distributions with means of 1 (``Small surround radius'' in Table \ref{tab:model}) or 3 (``Large surround radius''), both with standard deviation 0.3, and all other parameters the same as in our original simulation.  In separate simulations, we kept the mean surround radius at 2 but chose the relative surround strength from a Gaussian distribution with mean 0.5 (``Small surround weight'') or 2 (``Large surround weight''), both with standard deviation 0.1.  As with our original set of parameters, the models recovered from STRF analysis had at most slight biases toward weaker surrounds and shorter center times to peak under correlated noise (see Table \ref{tab:model}).

\bibliography{RetinaBibArXiv}

\clearpage

\section*{Supporting Information}
\makeatletter
\renewcommand{\thesubsubsection}{S\arabic{subsubsection}}
\makeatother
\setcounter{subsubsection}{0}
\makeatletter
\renewcommand\@seccntformat[1]{}
\makeatother

\subsubsection{Text S1. Dependence of output correlation on gain and firing rate.}\label{para:s1}  In the main text we found that gain adaptation contributes to decorrelation in a population of LN neurons. To gain further insight into why this should be the case, consider a pair of simplified LN neurons with logistic nonlinearity,
\begin{equation}
		P_i(spike | s) = 1 / \left( 1 + \exp\left( -g_i \left( s_i - b_i\right)\right)\right)
\end{equation}
and small firing probabilities in each time bin.  For neuron $i$, the gain of the model is $g_i$,  the linear filter output is $s_i$, and $b_i$ is an offset that will be adjusted to fix the average firing rate. Assuming a small firing probability amounts to assuming that $g_i (s_i  - b_i) \ll 0$ with high probability. Thus, the exponential term in $P_i(spike|s)$ dominates, and the stimulus-dependent firing probability simplifies to $P_i(spike | s) = \exp(g_i (s_i - b_i))$.  The average firing probability of one neuron is now
\begin{equation}
		P_i = \left<\exp(g_i (s_i - b_i))\right>,
\end{equation}
where the average is over the distribution of the filter output $s_i$, which can be approximated by the central limit theorem as a zero-mean Gaussian. (A nonzero mean could simply be absorbed into a redefinition of $b_i$.)  Using standard properties of Gaussian integrals, the averaging gives
\begin{equation}
	 	P_i = \exp\left[1/2 (g_i \sigma_i)^2 - g_ib_i\right],
\end{equation}
where $\sigma_i$  is the standard deviation of $s_i$. Note that, by the low firing probability assumption, $P_1 \ll 1$ and $P_2 \ll 1$. 

The average probability of simultaneous firing of two neurons is then given by
\begin{align}
		P_{12} &= \left<\exp(g_1 (s_1 - b_1))\exp(g_2 (s_2 - b_2))\right>\\
		     &= \left<\exp(g_1 s_1 +  g_2 s_2 -  g_1b_1 -  g_2b_2)\right>.
\end{align}
Assuming that the filter outputs are jointly Gaussian with correlation $\rho_s$, the variance of $g_1s_1 + g_2s_2$ is $(g_1\sigma_1)^2 + (g_2\sigma_2)^2 + 2(g_1 \sigma_1) \rho_s (g_2 \sigma_2)$ .  The expectation can therefore be computed as
\begin{align}
		P_{12} &= \exp\left[1/2((g_1\sigma_1)^2 + (g_2 \sigma_2)^2 + 2(g_1 \sigma_1) \rho_s (g_2 \sigma_2)) -  g_1b_1 -  g_2b_2\right]\\
	      &=  P_1 P_2 \exp\left[(g_1 \sigma_1) \rho_s (g_2 \sigma_2)\right].
\end{align}
The variance in firing of each neuron is $P_i(1-P_i)$, and the covariance between the two is $P_{12} - P_1P_2$.  The correlation coefficient of the two spike trains is then given by
\begin{equation}
		\rho =  \frac{P_{12} - P_1 P_2} {\sqrt{ P_1(1-P_1) P_2(1-P_2)}},
\end{equation}
which, using the above result for $P_{12}$ and taking the limit of small $P_i$, simplifies to 
\begin{equation}
		 \rho =  \sqrt{P_1 P_2} \left(e^{(g_1 \sigma_1) \rho_s (g_2 \sigma_2)} - 1\right).
\end{equation}
Thus there are three ways to reduce output correlations in this simple model: lower the overall firing rates, decrease $\rho_s$ by filter adaptation, or lower the rescaled gains $g_i \sigma_i$.

\subsubsection{Text S2. Tests of robustness.}\label{para:s2}  In our experiments, we alternated white noise and correlated stimuli.  The retina is known to adapt to a variety of stimulus features on timescales ranging from hundreds of milliseconds to a few seconds.  Thus, it is possible that estimating receptive fields using the entire trials confounds different states of adaptation.  To control for this possibility, we tested that our analyses were robust to leaving out the first 10 seconds of each trial.

In addition, we also varied the details of the analysis method, as summarized in Table S1.  The analysis method presented in the main text was used because it gave the most robustly unbiased results in our simulations, but we also wanted to verify that our results did not change dramatically with slightly different methods.  We first considered the possibility that we were including too many pixels in the surround.  To address this, we repeated our analysis but placed a threshold criterion on the surround so that only pixels positively correlated with the peak surround pixel were included.  Including only these pixels, center latency was still shorter for correlated noise than for white noise (Fig.\ \ref{f:latthresh}; ``Surround threshold'' in Table S1), while the increase in surround strength became more robust than in the original analysis (Fig.\ \ref{f:surrthresh}).  We then tested whether requiring the center to be contiguous was too stringent.  Removing this criterion did not change our overall results, although the center time to peak statistics are skewed by a few outliers (Fig.\ \ref{f:latcontig}, \ref{f:surrcontig}; ``Disconnected center''). 

When we computed STRFs, we included three frames after every spike so that we could measure any baseline offset in the estimated STRFs.  We generally subtracted the mean of these three frames from each STRF, but skipping this step did not affect our results (``No mean subtraction'').  As an additional check, we collapsed the full STRF into a single frame by projecting onto the first principal timecourse rather than the principal timecourse most similar to the peak pixel.  Making this change does not affect our results (``First principal component'').  (We obtained similar results when we chose the principal timecourse corresponding to the peak surround pixel rather than the peak center pixel.)

To investigate whether the changes we measured in receptive fields came from a change in the size or location of the receptive field center and surround or from a change in the receptive field strength at individual points in space, we repeated our standard analysis with the same masks for both stimuli.  That is, for each cell we first found the center region based on the STRF measured from white noise and then computed the time courses of this region under each stimulus condition from the full STRFs.  The surround time courses were computed similarly (``Masks from white noise'').  We then did the reverse, finding the center and surround regions from correlated noise STRFs and applying them to both stimuli (``Masks from correlated noise''). In either case, the center latencies were still larger for white noise, indicating that the time courses of individual pixels differ when stimulus correlations change.  On the other hand, the relative surround strength adaptation indices were centered around zero when masks were kept fixed.  Thus, any changes in surround strength observed in our main analysis were likely due to a subset of pixels switching from center to surround or vice versa.

\setcounter{figure}{0}
\makeatletter
\renewcommand{\thefigure}{S\arabic{figure}}
\makeatother
\begin{figure}
\begin{center}
\includegraphics[keepaspectratio, width=0.8\linewidth]{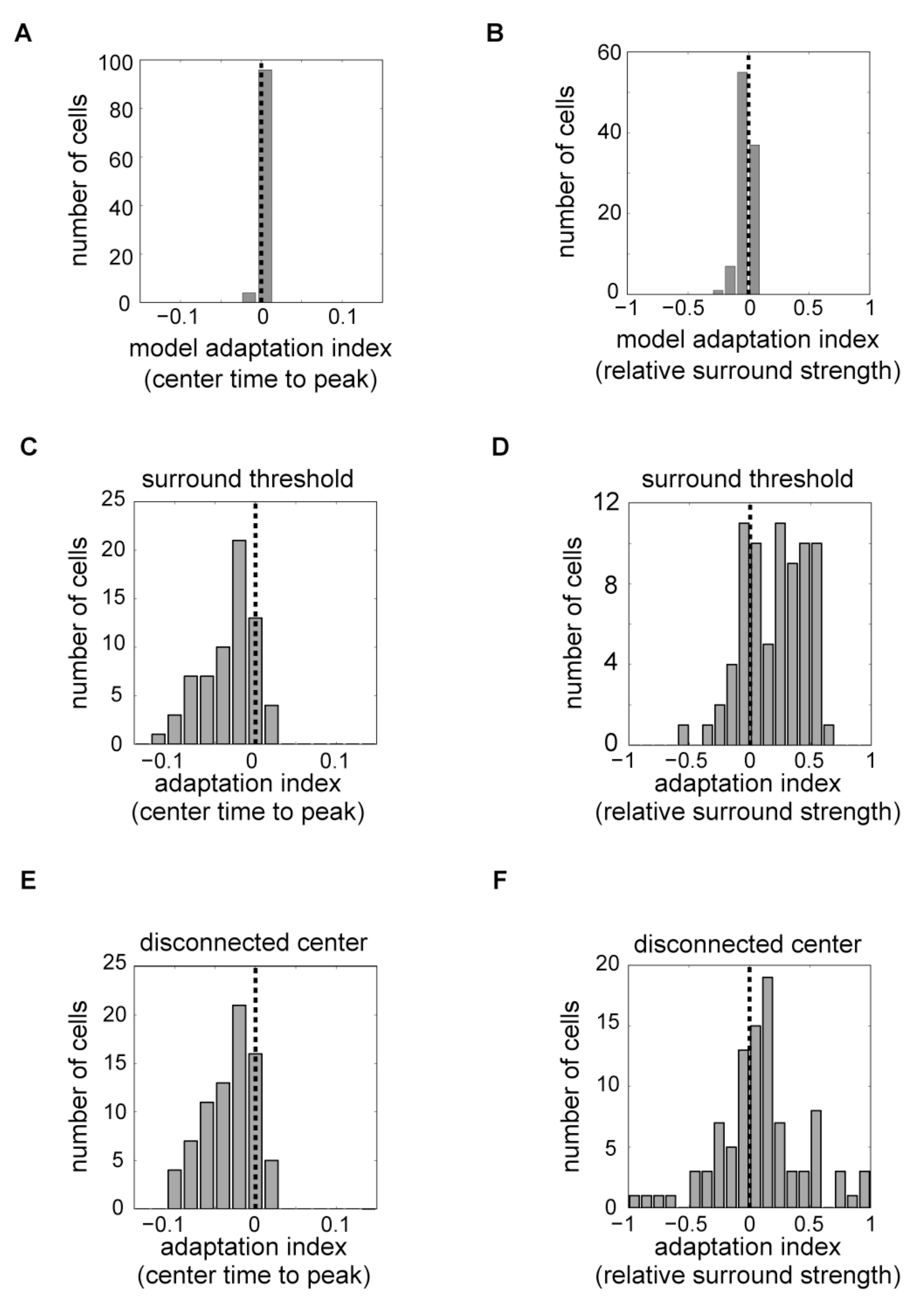}
\phantomsubcaption\label{f:latmodel}
\phantomsubcaption\label{f:surrmodel}
\phantomsubcaption\label{f:latthresh}
\phantomsubcaption\label{f:surrthresh}
\phantomsubcaption\label{f:latcontig}
\phantomsubcaption\label{f:surrcontig}
\end{center}
\caption{}
\label{f:FigS1}
\end{figure}

\begin{figure}\contcaption{
{\bf Receptive field results are validated with model neurons and are robust to analysis method changes.}  
\textbf{(A)} Center time to peak for a population of non-adapting model neurons, plotted as in Fig.\ \ref{f:latency}.  Receptive fields were constructed as a difference of Gaussians.  Surround radii (relative to center radii) had a mean of 2 and a standard deviation of 0.3.  Surround weights (relative to center weights) had a mean of 1 and a standard deviation of 0.1. The model neurons do not show a large difference between stimuli in center time to peak.
\textbf{(B)} Model neurons described in (A) show a slight bias toward smaller recovered relative surround strength under correlated noise compared to white noise. 
\textbf{(C)} Center time to peak is longer for white noise than for correlated noise when the surround only includes non-center pixels whose time courses are positively correlated with the time course of the peak negative pixel. 
\textbf{(D)} Relative surround strengths is marginally higher for correlated noise than for white noise under the same analysis as in (C). 
\textbf{(E)} Center time to peak is longer for white noise than for correlated noise when the center is not required to form a single contiguous component.  The figure omits a few outliers that lie outside the range of the horizontal axis and have longer time courses under correlated noise. 
\textbf{(F)} Relative surround strengths is marginally higher for correlated noise than for white noise under the same analysis as in (E). 
}
\end{figure}

\setcounter{table}{0}
\makeatletter
\renewcommand{\thetable}{S\arabic{table}}
\makeatother
\begin{table}
\caption{
\bf{Variants on receptive field analysis method.}}
{\small
\begin{center}\begin{tabular}{|l|c|c|c|c|c|c|c|c|}
\hline
& \multicolumn{4}{c|}{AI (relative surround strength)} & \multicolumn{4}{c|}{AI (center time to peak)}\\
& mean & std & $p$ & skew & mean & std & $p$ & skew\\
\hline
Standard analysis & 0.01 & 0.24 & .008 & 0.15 & $-0.03$ & 0.03 & $<.0001$ & $-0.53$\\
Disconnected center & 0.10 & 0.36 & .01 & $-0.01$ & $-0.03$ & 0.22 & .2 & $-2.81$\\
Surround threshold & 0.21 & 0.25 & $<.0001$ & $-0.40$ & $-0.03$ & 0.03 & $<.0001$ & $-0.59$\\
No mean subtraction & 0.12 & 0.23 & $<.0001$ & 0.08 & $-0.03$ & 0.03 & $<.0001$ & $-0.72$\\
First principal component & 0.07 & 0.24 & .01 & 0.11 & $-0.03$ & 0.03 & $<.0001$ & $-0.48$\\
Masks from WN & 0.02 & 0.20 & .4 & 0.04 & $-0.04$ & 0.04 & $<.0001$ & $-2.19$\\
Masks from CN & $-0.04$ & 0.19 & .08 & $-0.07$ & $-0.03$ & 0.03 & $<.0001$ & $-0.45$\\
\hline
\end{tabular}\end{center}
\begin{flushleft} Adaptation index (AI) in surround strength and center latency measured by variations in our analysis method.  Columns labeled ``mean,'' ``std,'' and ``skew'' show the mean, standard deviation, and skewness of the adaptation indices for the indicated analysis; columns labeled ``$p$'' show the $p$-values from $t$-tests of each distribution against the null hypothesis of zero mean. \textbf{Standard analysis}: Receptive fields were analyzed as presented in the main text. \textbf{Disconnected center}: The center was not required to form a contiguous region. \textbf{Surround threshold}: A threshold criterion was used to find the surround so that only pixels positively correlated with the peak surround pixel were included.  \textbf{No mean subtraction}: The mean of the frames after each spike was not subtracted from the STRFs. \textbf{First principal component}: The full STRF was collapsed onto a single frame by projecting onto the first principal timecourse rather than the principal timecourse most similar to the peak pixel. \textbf{Masks from WN}: Center and surround regions measured from white noise were used to analyze STRFs from both stimuli. \textbf{Masks from CN}: Center and surround regions measured from exponentially correlated noise were used to analyze STRFs from both stimuli.  
\end{flushleft}}
\label{tab:data}
\end{table}

\begin{figure}
\begin{center}
\includegraphics[keepaspectratio, width=\linewidth]{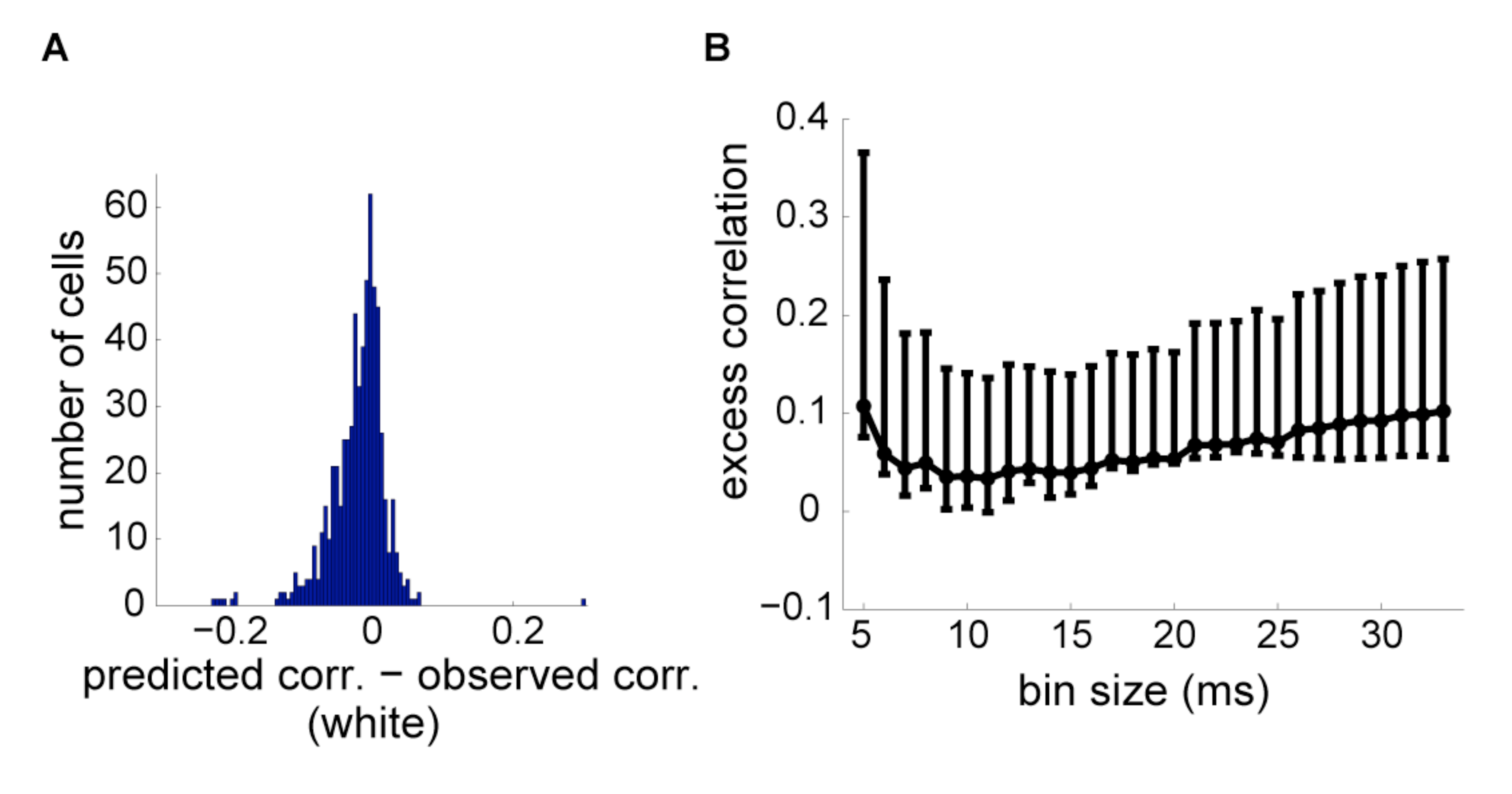}
\phantomsubcaption\label{f:corrvalidation}
\phantomsubcaption\label{f:binsize}
\end{center}
\caption{
{\bf Correlation measurement controls.}  
\textbf{(A)} As a control on the quality of the non-adapting LN model, we examined the difference between its predicted pairwise correlations under the white noise stimulus (the stimulus to which the model was fit) and the observed correlations. Since the model is a single-neuron model that does not attempt to capture pairwise correlations, we do not expect it to reproduce these correlations perfectly. Nevertheless, many cell pairs are well-predicted, indicating that their correlation is largely due to receptive field overlap. There is, however, a slight tendency for the model to underestimate correlations: this is likely due to its neglect of noise correlations. We note that such a bias will not effect the model's predicted \emph{excess correlation}, unless it is very different in the two stimulus conditions. But such an effect would represent a form of non-trivial processing in its own right. 
\textbf{(B)} Our correlation measurements were based on binned spike trains. We measured excess correlation, in the spatiotemporal exponential dataset, for a variety of bin sizes. Its value is largely independent of bin size. Error bars represent $95\%$ bootstrap confidence intervals. All correlations reported in the main text were estimated from spike trains binned at $33$ ms. 
}
\label{f:FigS2}
\end{figure}

\end{document}